\documentclass[,final]{aipproc}
\layoutstyle{6x9}

\begin{document}

\title{Slepton pair production at hadron colliders}



 \keywords{}

\author{Benjamin Fuks}{address={Laboratoire de Physique Subatomique et de
Cosmologie,\\ Universit\'e Joseph Fourier/CNRS-IN2P3, 53 Avenue
des Martyrs, F-38026 Grenoble, France}}

\begin{abstract}
In R-parity conserving supersymmetric (SUSY) models, sleptons are
produced in pairs at hadron colliders. We show that measurements
of the longitudinal single-spin asymmetry at possible polarization
upgrades of existing colliders allow for a direct extraction of
the slepton mixing angle. A calculation of the transverse-momentum
($q_T$) spectrum shows the importance of resummed contributions at
next-to-leading logarithmic (NLL) accuracy in the small and
intermediate $q_T$-regions and little dependence on unphysical
scales and non-perturbative (NP) contributions.

\end{abstract}

\maketitle



\section{Introduction}

The Minimal Supersymmetric Standard Model \cite{Nilles:1983ge,
Haber:1984rc}, one of the most promising extensions of the
Standard Model (SM), postulates a symmetry between fermionic and
bosonic degrees of freedom in nature, predicting thus the
existence of a fermionic (bosonic) SUSY partner for each bosonic
(fermionic) SM particle. Its main advantages are the stabilization
of the gap between the Planck and the electroweak scale
\cite{Witten:1981nf}, gauge coupling unification at high energy
scales \cite{Dimopoulos:1981yj}, and a stable lightest
supersymmetric particle as a dark matter candidate
\cite{Ellis:1983ew}. Spin partners of the SM particles have not
yet been observed and in order to remain a viable solution to the
hierarchy problem, SUSY must be broken at low energy via soft mass
terms in the Lagrangian. As a consequence, the SUSY particles must
be massive in comparison to their SM counterparts, and the
Tevatron and the LHC will perform a conclusive search covering a
wide range of masses up to the TeV scale.

We focus on slepton pair (slepton-sneutrino associated) production
at hadron colliders through Drell-Yan type processes $q\,\bar{q}
\to \gamma,\,Z^0 \to \tilde{l}_i\, \tilde{l}_j^\ast$, and $q\,
\bar{q}^\prime \to W^\mp \to \tilde{l}_i\, \tilde{\nu}_l^\ast,\,
\tilde{l}_i^\ast\, \tilde{\nu}_l~$. Due to their purely
electroweak couplings, sleptons are among the lightest SUSY
particles in many SUSY breaking scenarios and often decay directly
into the stable lightest SUSY particle plus the corresponding SM
partner. A slepton-pair signal at hadron colliders will therefore
consist in a highly energetic lepton pair, which will be easily
detectable, and associated missing energy.


\section{Fixed order calculations}

\begin{figure}
 \includegraphics[height=.28\textheight]{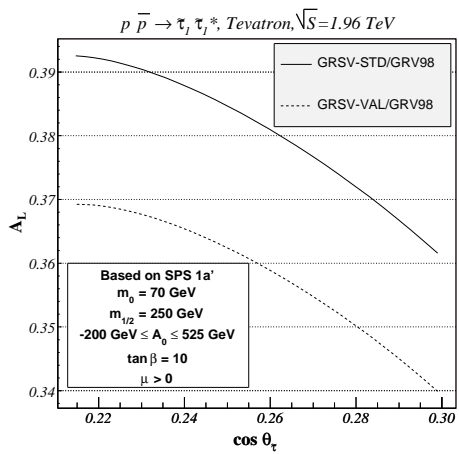}
 \caption{\label{fig:asym}Dependence of the longitudinal
  single-spin asymmetry on the cosine of the stau mixing angle for
  $\tilde{\tau}_1$ pair production in a typical mSUGRA model at
  the Tevatron.}
\end{figure}

The neutral current cross section for the production of non-mixing
slepton pairs in collisions of quarks with definite helicities has
been calculated in \cite{Chiappetta:1985ku}, and an extension
including the mixing of the left- and right-handed interaction
eigenstates has been performed in \cite{Bozzi:2004qq}.
Experimentally, proton beams are much more easily polarized than
antiproton beams, and it may be easy to implement the polarization
of the proton beam for a possible upgrade of the Tevatron
\cite{Baiod:1995eu}.

In Fig.\ \ref{fig:asym}, we show the longitudinal single-spin
asymmetry $A_L$ as a function of the cosine of the stau mixing
angle, for a typical mSUGRA scenario based on the SPS 1a' model
\cite{Aguilar-Saavedra:2005pw} where the trilinear coupling $A_0$
varies. The predictions are very sizeable in the entire viable
$A_0$ range and depend strongly on the stau mixing angle.
Unfortunately, the parton density uncertainty is still large, but
it should be reduced considerably in the future through more
precise measurements. Furthermore, the SM background can be
distinguished from the SUSY signal through its asymmetry with
opposite sign.

\begin{figure}
 \includegraphics[height=.22\textheight]{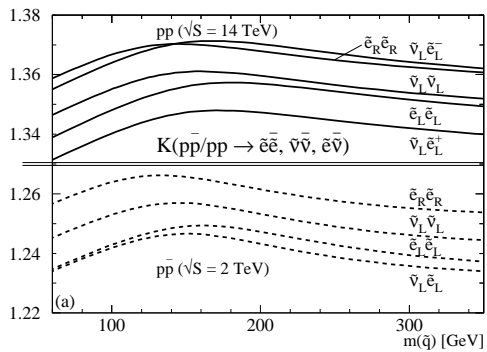}
 \caption{\label{fig:NLO}Ratios of the total next-to-leading order
 (NLO) cross sections to the corresponding LO cross
 sections for slepton-pair production at the LHC (full) and
 Tevatron (dashed) as a function of the squark mass
 $m_{\tilde{q}}$ for a gluino mass of 200~GeV and slepton masses
 of ${\tilde{e}_R}/{\tilde{e}_L}/{\tilde{\nu}_L} = 120/150/135$
 GeV.}
\end{figure}

The QCD corrections to the total slepton-pair hadroproduction
cross section have been calculated in \cite{Baer:1997nh}, and a
complete analysis including the SUSY-QCD corrections has been
performed in \cite{Beenakker:1999xh}. Since massive squarks and
gluinos are involved in the loops, the genuine SUSY corrections
are expected to be considerably smaller than the standard QCD
ones, as it is presented in Fig.\ \ref{fig:NLO}. The SUSY-QCD
$K$-factors are approached in the asymptotic limit of large
$\tilde{q}/\tilde{g}$ masses by the QCD $K$-factors.

\section{Transverse-momentum spectrum at NLL}

A precise knowledge of the $q_T$-balance is of vital importance
for the discovery of SUSY particles. The Cambridge (s)transverse
mass proves to be particularly useful for the determination of
slepton masses \cite{Lester:1999tx} and spin \cite{Barr:2005dz},
the two key features distinguishing them from SM leptons
\cite{Lytken:22, Andreev:2004qq}. Furthermore, both detector
kinematical acceptance and efficiency depend on $q_T$.

When studying the $q_T$-distribution of a slepton pair produced
with invariant mass $M$ in a hadronic collision, it is appropriate
to separate the large- and small-$q_{T}$ regions. At large $q_T$,
the use of fixed-order perturbation theory is fully justified,
since the perturbative series is controlled by a small expansion
parameter, $\alpha_s(M^2)$. However, at small $q_{T}$, the
coefficients of the perturbative expansion are enhanced by powers
of large logarithmic terms, $\ln(M^{2}/q_{T}^{2})$, and
fixed-order calculations diverge as $q_T \to 0$. These logarithms
are due to multiple soft-gluon emission from the initial state and
have to be resummed to all orders in $\alpha_s$ in order to obtain
reliable predictions. At intermediate $q_T$, the resummed result
has to be consistently matched with fixed-order perturbation
theory in order to obtain predictions with uniform theoretical
accuracy over the entire $q_T$-range.

\begin{figure}
 \includegraphics[scale=0.39]{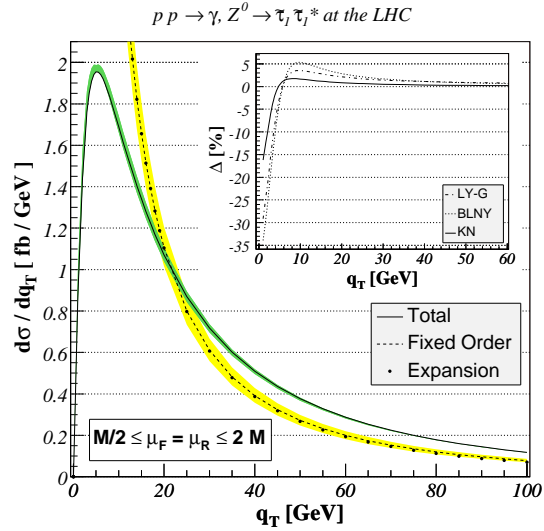}
 \caption{\label{fig:res} Differential cross section for the
 process $p p \to \tilde{\tau}_{1}\tilde{\tau}_{1}^{*}$ at the
 LHC. NLL+LO matched result, LO result, asymptotic expansion of
 the resummation formula and $\Delta$-parameter are shown.}
\end{figure}

We have recently implemented the formalism proposed in
\cite{Bozzi:2005wk} and computed the $q_T$-spectrum of a slepton
pair produced at the LHC by combining NLL resummation and ${\cal
O}(\alpha_s)$ perturbation theory in \cite{Bozzi:2006fw}. In Fig.\
\ref{fig:res}, we choose the SPS7 mSUGRA benchmark point
\cite{Allanach:2002nj}, which gives a light ${\tilde \tau}_{1}$ of
114 GeV, and we show the $q_T$-distribution for $\tilde{\tau}_1$
pair production. We can see that the LO result (dashed line)
diverges as $q_{T}\to 0$, and the asymptotic expansion of the
resummation formula at ${\cal O}(\alpha_s)$ (dotted line) is in
very good agreement with LO at small and intermediate $q_{T}$. In
the small and intermediate $q_T$-regions, the effect of
resummation (solid line) is clearly visible.

The quantity $\Delta$ gives an estimate of the contributions from
different NP parameterizations (LY-G \cite{Ladinsky:1993zn}, BLNY
\cite{Landry:2002ix}, KN \cite{Konychev:2005iy}), which are under
good control since their effect is always less than 5\% for $q_{T}
>$5 GeV. The perturbative uncertainty, estimated by allowing
$\mu_{F}=\mu_{R}$ to vary between $M/2$ and $2M$, is clearly
improved with respect to the pure fixed-order calculations,
reducing the scale dependence from 10\% to 5\% for $q_{T}$ values
up to 100 GeV. Moreover, when integrated over $q_T$, the NLL+LO
curve leads to a total cross section in good agreement with the
total cross section at ${\cal O}(\alpha_s)$
\cite{Beenakker:1999xh}.

\section{Conclusions}

The recent luminosity upgrade of the Tevatron and the imminent
start-up of the LHC put the discovery of SUSY particles with
masses up to the TeV-scale within reach. Since this discovery
depends critically on the missing transverse-energy signal, a
precise knowledge not only of the total cross section, but also of
the $q_T$-spectrum is mandatory. We have performed a first step in
this direction by resumming multiple soft-gluon emission for
slepton-pair production at NLL level and matching it to the
fixed-order calculation. (S)transverse mass measurements will then
lead to precise slepton mass and spin determinations, while
polarization of the initial hadron beams may allow for an
extraction of the slepton mixing angle and the underlying
SUSY-breaking parameters.


\bibliographystyle{aipproc}   

\end{document}